# Specific Heat of $Na_{0.35}CoO_2 \cdot 1.3H_2O$: Effects of Sample Age; Non-Magnetic Pair Breaking, Two Energy Gaps, and Strong Fluctuations in the Superconducting State

N. Oeschler,[1] R. A. Fisher,[1] N. E. Phillips,[1] J. E. Gordon,[2] M.-L. Foo,[3] and R. J. Cava[3]

[1]*Lawrence Berkeley National Laboratory and Department of Chemistry, University of California, Berkeley, CA 94720*
[2]*Physics Department, Amherst College, Amherst, MA 01002*
[3]*Department of Chemistry, Princeton University, Princeton, NJ 08544*

The specific heats of three samples of $Na_{0.35}CoO_2 \cdot 1.3H_2O$ show an evolution of the superconductivity, and its ultimate disappearance, with increasing sample age. An overall increase in pair-breaking action, which occurs preferentially in an electron band with a small energy gap, produces a shift in the relative contributions of two electron bands to the superconducting condensation. The similarity of the time scale for these changes to that recently reported for structural changes in the $CoO_2$ layers and the formation of O vacancies suggests a relation between the two effects and an explanation for the strong sample dependence of the properties of this material more generally. The onset of the transition to the vortex state is independent of magnetic field, suggesting the presence of unusually strong fluctuation effects.

The discovery of superconductivity in the cuprates, with critical temperatures ($T_c$) as high as 133 K, raised the question of whether high-$T_c$ superconductivity might be found in similar systems with other ions replacing the Cu. Although Co was recognized as an interesting candidate almost immediately, the very fragile superconductivity of $Na_{0.35}CoO_2 \cdot 1.3H_2O$, with $T_c$ ~4.5 K, was not discovered until 2003 [1]. Most of the structural and electronic features thought to be important for the superconductivity in the cuprates occur in $Na_{0.35}CoO_2 \cdot 1.3H_2O$, but in the cuprates Cu ions in an approximately *square* array are ordered antiferromagnetically, and spin fluctuations are thought to play a role in the electron pairing, while in $Na_{0.35}CoO_2 \cdot 1.3H_2O$ Co ions in a *triangular* array are magnetically frustrated, which may affect the superconductivity. Theoretical work suggests that the superconductivity is different from that of the cuprates, and possibly unique. The conduction-electron contribution to the superconducting-state specific heat ($C_{es}$) gives information about the symmetry of the order parameter (OP) and clues to the nature of the pairing. A number of specific-heat measurements have been reported [2-5], but in many cases the interpretation of the results is limited by contributions from paramagnetic centers, lack of data at sufficiently low temperatures, or apparent experimental error. The sample-to-sample variation of the other results has prevented an unambiguous identification of the relevant intrinsic properties. In general, the sample dependence of the properties of this material has been assumed to be a consequence of unidentified variations in "sample quality". In the case of the specific heat, and in analogy with early measurements on the cuprates, it has been attributed to incomplete transitions to the superconducting state, and the presence of variable amounts of "normal material". Our results on two superconducting samples show that this interpretation is incorrect. Instead, they are consistent with recent evidence of inherent structural changes related to sample age, which suggest that the two samples have to be considered as two different, but closely related, superconductors.

We have measured the specific heat ($C$) of three polycrystalline samples of $Na_{0.35}CoO_2 \cdot 1.3H_2O$. They were prepared by the same procedure [6], and differed only in the time they were kept at room temperature and 100% relative humidity before the measurements, approximately 40, 3, and 5 days, for Samples 1, 2, and 3, respectively. Measurements on Sample 1, which showed a weakly field-dependent specific-heat anomaly near 7 K, possibly a CDW transition, but no superconductivity, will be reported elsewhere. Measurements on Samples 2 and 3, with $T_c$ ~4.5 K and no significant contribution from paramagnetic centers, are reported here. They extend over the temperature ($T$) intervals 0.85 to 32 K, and 0.34 to 14 K, respectively, and include measurements in magnetic fields ($B$) to 9 T for Sample 2. In zero field there are substantial normal-state-like contributions to $C$, $\gamma_r T$, that correspond to a "residual" electron density of states (DOS). A similar, sample-dependent, residual DOS is seen in every measurement that permits an extrapolation to 0 K, and also in the nuclear-spin relaxation time ($T_1$) [7,8]. The sample-dependence of $\gamma_r$ has been interpreted in terms of differences in the "volume fraction of superconductivity", with the assumption that the superconducting phase is the same. Differences in $C_{es}$ for Samples 2 and 3 are not consistent with that assumption. They are consistent with recent evidence that the superconducting forms of this material are thermodynamically unstable and the superconductivity evolves with sample age on a time scale of several days [9,10], and finally disappears. Changes in $T_c$ are accompanied by changes in the residual DOS [9], Co oxidation state, crystal structure, and concentration of O vacancies [10]. In this context, the variations of $C_e$ and $\gamma_r$, for all three of our samples, correspond to a progression through a series of closely related superconducting materials that terminates in a nonsuperconducting material. The properties measured *are* intrinsic, but correspond to slightly different superconductors. The measurements reported here give information on the effect on $C_{es}$ of the underlying structural and electronic changes. This interpretation of the sample dependence also emphasizes the difficulties inherent in any detailed comparison of experiment with theory for this interesting but exceptionally complicated material.

The general nature of the results is illustrated by $C(0)$ and $C(9T)$ for Sample 2 in Figs. 1(a) and 1(b), which show the specific-heat anomaly at $T_c$ and the absence of other transitions below 30 K. A $T^{-2}$ term, the low-$T$ "upturn" in the 9-T data in Fig. 1(a), and similar terms in 3, 5, and 7 T, are in good agreement with calculated hyperfine contributions ($C_{hyp}$). After subtraction of the $T^{-2}$ term, $C(9T)$ to 12 K was combined with $C(0)$ from 6 to 12 K, and fitted as $C = C_{en} + C_{lat}$, where $C_{en} = \gamma_n T$ is the normal-state electron contribution, and $C_{lat} = B_3 T^3 + B_5 T^5 + B_7 T^7$ is the lattice contribution. The fit gave $\gamma_n = 16.1$ mJ K$^{-2}$ mol$^{-1}$, $B_3 = 0.126$ mJ K$^{-4}$ mol$^{-1}$, and $T_c = 4.52$ K. For $B \leq 1$ T, small $T^{-2}$ terms, *e.g.*, an upturn in the $B = 0$ data, that is barely perceptible in Fig. 1(a), are evidence of a "magnetic" contribution ($C_{mag}$) associated with paramagnetic centers at a concentration of ~$10^{-3}$ mol/mol sample, which make no observable contribution to $C(B)$ at higher $T$ (see Figs. 4 and 1(c)). The conduction-electron contribution is $C_e(B) = C(B) - C_{lat} - C_{hyp}(B) - C_{mag}(B)$; for $T < T_c$, $C_e(0) = C_{es}$. In 9 T there is no indication of the transition to the vortex state that is apparent for all other fields (see Fig. 4), showing that $B_{c2}$ is ~9 T, and $C_{en} = C_e(9T)$. $C_{es}$ and $C_{en}$ are shown in Fig. 1(c). Below 2 K, $C_{es}/T$ is linear in $T$, and extrapolation to $T = 0$ gives a 6.5-K entropy that agrees with those calculated from the in-field data (to within 1% for 9 T, and ± 0.5% for other fields). The 0-K intercept is $C_e(0)/T \equiv \gamma_r = 6.67$ mJ K$^{-2}$ mol$^{-1}$; the $T^2$ term, a signature of line nodes in the energy gap (but see below), has also been identified by Yang *et al.* [3] in another sample (Sample A in the following).

For Sample 3 there is no evidence of paramagnetic impurities. Fitting $C(0)$ with the expressions and in the $T$ interval used for Sample 2, and with the constraint that the normal- and superconducting-state entropies be equal at 6.5 K, gave $\gamma_n = 15.7$ mJ K$^{-2}$ mol$^{-1}$ and $T_c = 4.65$ K. There is no evidence of a $T^2$ term in $C(0)$: the lowest-$T$ data can be fitted as the sum of an exponential term, expected for a "fully-gapped" superconductor, and $\gamma_r T$, with $\gamma_r = 11.0$ mJ K$^{-2}$ mol$^{-1}$.

Two mechanisms have been invoked to explain the residual DOS: pair breaking by scattering centers, and the presence of normal material associated with defects that suppress the superconductivity in regions on the scale of the coherence volume but do not affect the superconductivity elsewhere. The specific-heat results have been interpreted in terms of normal material. It is assumed that the superconductivity occurs in a fraction, $(\gamma_n - \gamma_r)/\gamma_n$, of the sample, and $C_e(B)$ for one mole of superconducting material becomes $C_e'(B) \equiv [C_e(B) - \gamma_r T][\gamma_n/(\gamma_n - \gamma_r)]$. $C_e'(0)$, which is $C_{es}'$ for $T \leq T_c$, is shown in Figs. 2 and 3 for Samples 2 and 3. Measurements on a single sample would not distinguish between pair breaking and normal material, but the marked differences between $C_{es}'$ for Samples 2 and 3 shows that the superconductivity is *different* in the two samples, and the normal-material model is not applicable. The measurements of $T_1$ have been interpreted in terms of pair breaking. Since the same nuclei see both the change in DOS at $T_c$ and the residual DOS, they do distinguish between normal material and pair breaking, and require the latter. With this interpretation of $\gamma_r$, $C_{es}'$ would be only an approximation to $C_{es}$ for the "pure" superconducting state, but there is reason to expect it to show the main features: The dependences of $T_1$ and $C_{es}$ on the DOS are closely related; the $T_1$ data show a fairly sharp transition from the $T$ dependence associated with the residual DOS at low $T$ to that characteristic of the pure superconducting state at higher $T$ [8]; calculations for a particular example of pair breaking [11] account for the behavior of $T_1$ and give a DOS that is qualitatively consistent with $C_{es}'$. In the following, we use this approximation as a basis for an interpretation of $C_{es}$.

The unusual deviations from BCS theory for Sample 2 (see Fig. 2) are similar to those for MgB$_2$ [4], which suggests the presence of two energy gaps, the origin of the deviations for MgB$_2$. Penetration-depth measurements have also been interpreted in terms of two gaps [12]. Several two-gap fits, in which $C_{es}'$ is the sum of contributions characterized by the 0-K energy-gap parameters, $\Delta_i(0)$, represented by $\alpha_i \equiv \Delta_i(0)/k_B T_c$, and the fractional contributions to the normal-state density of states, $\gamma_i/\gamma_n$, $i = 1, 2$, are shown in Fig. 2. The dotted curves represent a fit based on the assumption of line nodes in the small gap, with $\alpha_1 = 2.15$, $\alpha_2 = 1.00$, $\gamma_1/\gamma_n = 0.55$, and $\gamma_2/\gamma_n = 0.45$. The small-gap contribution is calculated using the BCS temperature dependence for the gap, but allowing for the presence of line nodes to reproduce the $T^2$ behavior observed for $0.2 \leq$ $T/T_c \leq 0.4$, and extend it to 0 K. If there are nodes, they might be expected to be common to all bands, but $Sr_2RuO_4$ appears to be an exception, and the contribution of the small-gap band is very similar to that of one of the three bands in $Sr_2RuO_4$ [13]. The absence of evidence for nodes in Sample 3 led to an alternative interpretation of the Sample-2 data based on a "fully-gapped" model: The solid curves in Fig. 2 represent a two-gap fit, with $\alpha_1 = 2.20$, $\alpha_2 = 0.70$, $\gamma_1/\gamma_n = 0.55$, $\gamma_2/\gamma_n = 0.45$, without nodes. With these parameters, the *observed* $T^2$ dependence is reproduced without nodes. The breadth of the transition is no doubt partly a consequence of inhomogeneity in the sample, but there is also reason to think that fluctuations are important (see below). These two effects cannot be reliably separated, but two possibilities are represented in the inset to Fig. 2.

The deviations from BCS theory are not as great for Sample 3 as for Sample 2, but they still suggest two gaps. Near $T_c$, the high values of $C_{es}'$ and positive curvature of $C_{es}'/T$ are signatures of strong coupling, *i.e.*, $\alpha > 1.764$, the BCS weak-coupling value. The best single-gap fit, which approximates the data near $T_c$, underestimates $C_{es}'$ at low temperatures, where the small-gap contribution is important. A two-gap model without nodes, with $\alpha_1 = 2.30$, $\alpha_2 = 1.10$, $\gamma_1/\gamma_n = 0.80$, and $\gamma_2/\gamma_n = 0.20$, gives a good fit.

For Sample 3 the pair-breaking action, which is measured by the value of $\gamma_r$, is substantially stronger than for Sample 2, and the $T$ dependence of $C_{es}'$ is qualitatively different. The low-$T$ $T^2$ dependence of $C_{es}'$ in Sample 2 is replaced by an exponential dependence, and the strong-coupling effects, as measured by $\Delta C_e'(T_c)/\gamma_n T_c$, are more pronounced. The suggestion that the properties of Samples 2 and 3 are "intrinsic", and the differences between Samples 2 and 3 are not a consequence of random differences in "sample quality", is supported by the similarities between the properties of Samples 2 and 3 and those of another pair of samples, Sample A and Sample B, another sample studied by Yang et al. [5]. These similarities include the values of $T_c$, the general temperature dependences of $C_{es}'$ and the correlation of those temperature dependences with $\gamma_r/\gamma_n$: For Sample A, $\gamma_r/\gamma_n = 0.53$, $\Delta C_e'(T_c)/\gamma_n T_c = 1.45$, and $T_c = 4.5$ K (*cf* 0.41, 1.35, and 4.52 K for Sample 2); $C_e/T$ *vs* $T$ is close to linear for $1 < T < 3$ K, just as it is for Sample 2. For Sample B, $\gamma_r/\gamma_n = 0.73$, $\Delta C_e'(T_c)/\gamma_n T_c = 1.96$, and $T_c = 4.7$ K (*cf* 0.70, 2.08, and 4.65 K for Sample 3); $C_e/T$ *vs* $T$ shows substantial positive curvature for $1 < T < 3$ K, just as it does for Sample 2.

Mixed-state $C_e'(B)$, data for Sample 2 are shown in Fig. 4. The temperature of the onset of the transition to the mixed state is independent of $B$. This effect is expected for strong fluctuations, but it is unusually large. Values of $\gamma(B)$, the intercepts obtained by the linear extrapolations to 0 K, are shown in the inset. The "shoulder" near 3 T is at least qualitatively consistent with the anisotropy of $B_{c2}$ [14]. The $B^{1/2}$ dependence and the sharp increase in $\gamma(B)$ at low field reported in Ref. 3 are related to a low-$T$ curvature of $C_e(B)/T$ vs $T$, which is not present in the data reported here. Neither the results presented here nor those of Ref. 3 extend into the field and temperature region in which the $B^{1/2}$ dependence is expected for line nodes [15].

The differences between Samples 2 and 3 are related to different sample ages, 5 and 3 days, and it is reasonable to associate them with structural and electronic changes that take place on a similar time scale [10]. Identification of the pair-breaking scattering centers with O vacancies, which have been shown to have a concentration ($\delta$) that depends on room-temperature sample age, is consistent with the relevant experimental results, and provides a possible explanation of the differences between our samples. As a function of increasing sample age, $T_c$ goes through a broad maximum, ~4.5 K, that separates non-superconducting regions, and is accompanied by a regular increase in δ and changes in the Co oxidation state [10]. The time scales depend on storage conditions [9,10] but it is reasonable to assume that Sample 3 has a higher value of δ than Sample 2, but essentially the same $T_c$, while Sample 1, with a still higher δ, would not be superconducting. This would account for the increased pair breaking in Sample 3, with the decrease in $T_c$ that would normally be expected to accompany the increased pair breaking compensated by other changes in electronic structure. This interpretation, and the assumption that the difference in $C_{es}'$ for Samples 2 and 3 is a consequence of the pair breaking implies correlations among $\gamma_r$, $C_{es}'$, and $T_c$ that have implications for possible values of $\gamma_r$: High values of $T_c$ are associated with substantial values of δ [10], which suggests that *any* sample with a high $T_c$ will also have a substantial $\gamma_r$. This correlation suggests that, unlike the cuprates and heavy-Fermion superconductors, improvements in "sample quality" may not reduce $\gamma_r$. It is also noteworthy that the increased pair breaking in Sample 3 occurs preferentially in the small-gap band, in which the pairing interaction is weaker: The values of $\gamma_i/\gamma_n$ were obtained by multiplying $C_e$ by $\gamma_n/(\gamma_n - \gamma_r)$. The real fractional contributions of the DOS to the superconducting transition, obtained by dividing by those factors, are 0.32 and 0.26, respectively, for the large- and small-gap bands in Sample 2, and 0.24 and 0.06 in Sample 3.

Pair breaking by non-magnetic scattering centers rules out an isotropic s-wave OP. Among the many OP that have been considered, there is no experimental evidence for the triplet-spin pairing that would correspond to a p-wave OP, and we confine our consideration to a d-wave OP, the $d_{x2-y2}$ state with nodes, and the nodeless $d_{x2-y2}+id_{xy}$ state. The evidence for nodes is ambiguous: $C_{es}'$ for Sample 2 can be fitted with or without nodes. Data for Sample A [3] also show the $T^2$ behavior characteristic of nodes and extend to slightly lower $T$, but they could also be fitted with an expression for two-gaps without nodes. For Sample 3 there is no evidence of nodes, but that could be a consequence of the strong pair breaking in that band filling the low-energy states. Data for Sample B [5] were approximated by a $T^2$ term in a small interval of $T$, but they are very similar to those for Sample 3, and it is reasonable to assume that they would show the same exponential dependence if they extended to lower $T$. Data for $T_1$ have been interpreted as showing the presence of line nodes [7,8], but the power-law $T$ dependences on which those conclusions were based were not observed at low $T$, where they are expected theoretically, but near $T_c$, where they could be manifestations of the strong-coupling effects that are apparent in $C_{es}'$. This leaves comparisons with calculations as a basis for deciding between OP with and without nodes: Calculations [11] for unitary scattering with the $d_{x2-y2}+id_{xy}$ OP give better agreement with the $T_1$ data [8]. The same calculation gives a DOS that is qualitatively consistent with the clean separation of $C_{es}$ for Sample 3 into the $\gamma_r T$ and exponential terms; with a $d_{x2-y2}$ OP it would give a strong $T^2$ dependence that is not observed. We conclude that the $d_{x2-y2}+id_{xy}$ OP, which accounts for the $T_1$ data, and which has been predicted on the basis of a t-J model [16], is also the most consistent with the $C_{es}'$ data.

LDA and LSDA band-structure calculations [17,18] give six small hole pockets near the K points as well as a cylindrical surface around the Γ point. However, the hole pockets are sensitive to the doping level and to the on-site Coulomb interaction (U), the magnitude of which is not well known, and disappear for U ~2 eV [18]. The specific heat and the penetration depth [12] give evidence of two energy gaps, the existence of which would be difficult to understand without distinctly different sheets of the Fermi surface. They are, therefore, more consistent with the existence of the pockets, which are of particular interest in connection with a theory [19] in which the electron pairing is mediated by spin fluctuations associated with their nesting character.

We thank B. H. Brandow, D.-H. Lee, I. I. Mazin, D. J. Singh, P. Zhang, V. Kresin, and R. E. Walstedt for helpful comments and discussions. The work at LBNL was supported by the Director, Office of Basic Energy Sciences, Materials Sciences Division of the U. S. DOE under Contract No. DE–AC03–76SF00098; at Princeton, by NSF grant DMR-0213706 and a DOE-BES grant DE-FG02-98-ER45706. N. O. was supported, in part, by the DAAD.

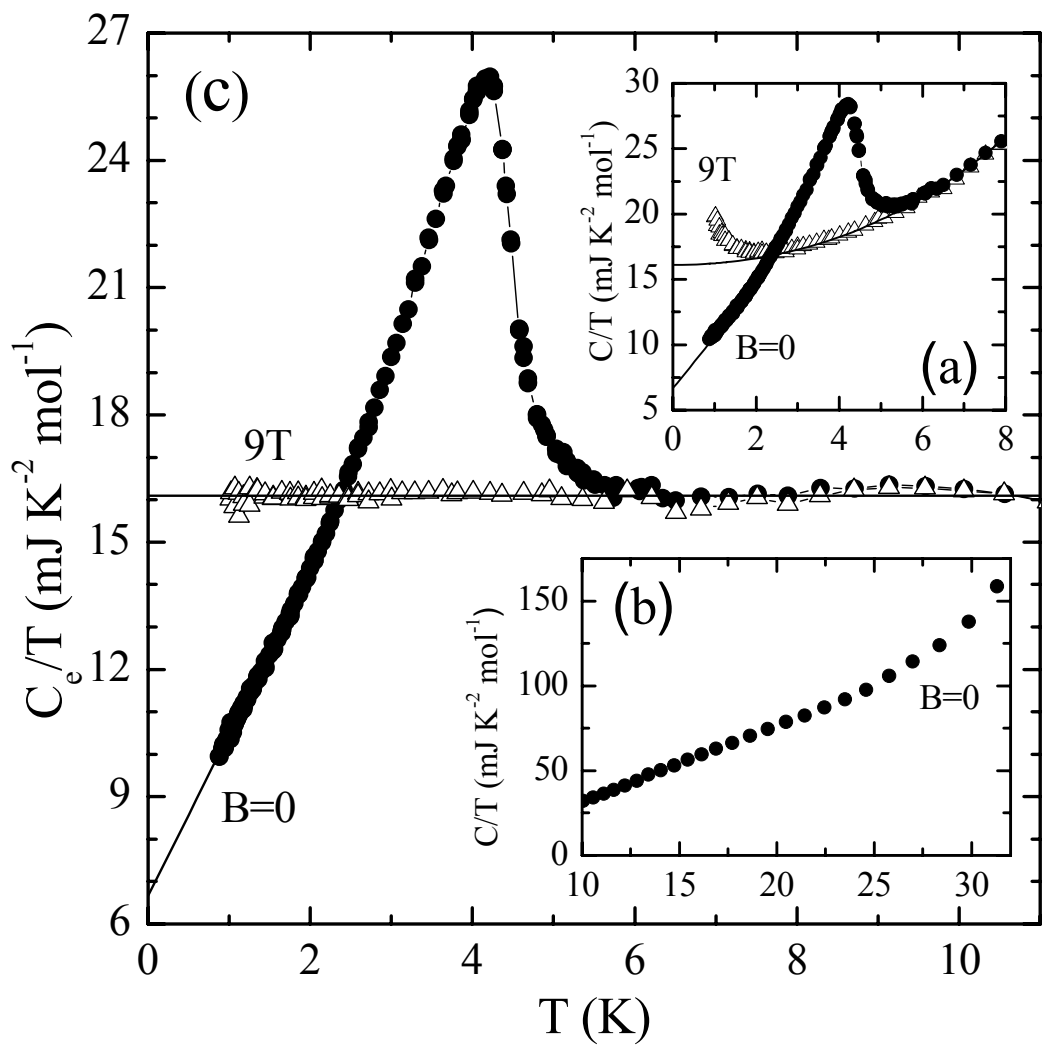

FIG. 1. Sample-2 specific-heat: (a,b) the total specific heat; (c) the electron contribution.

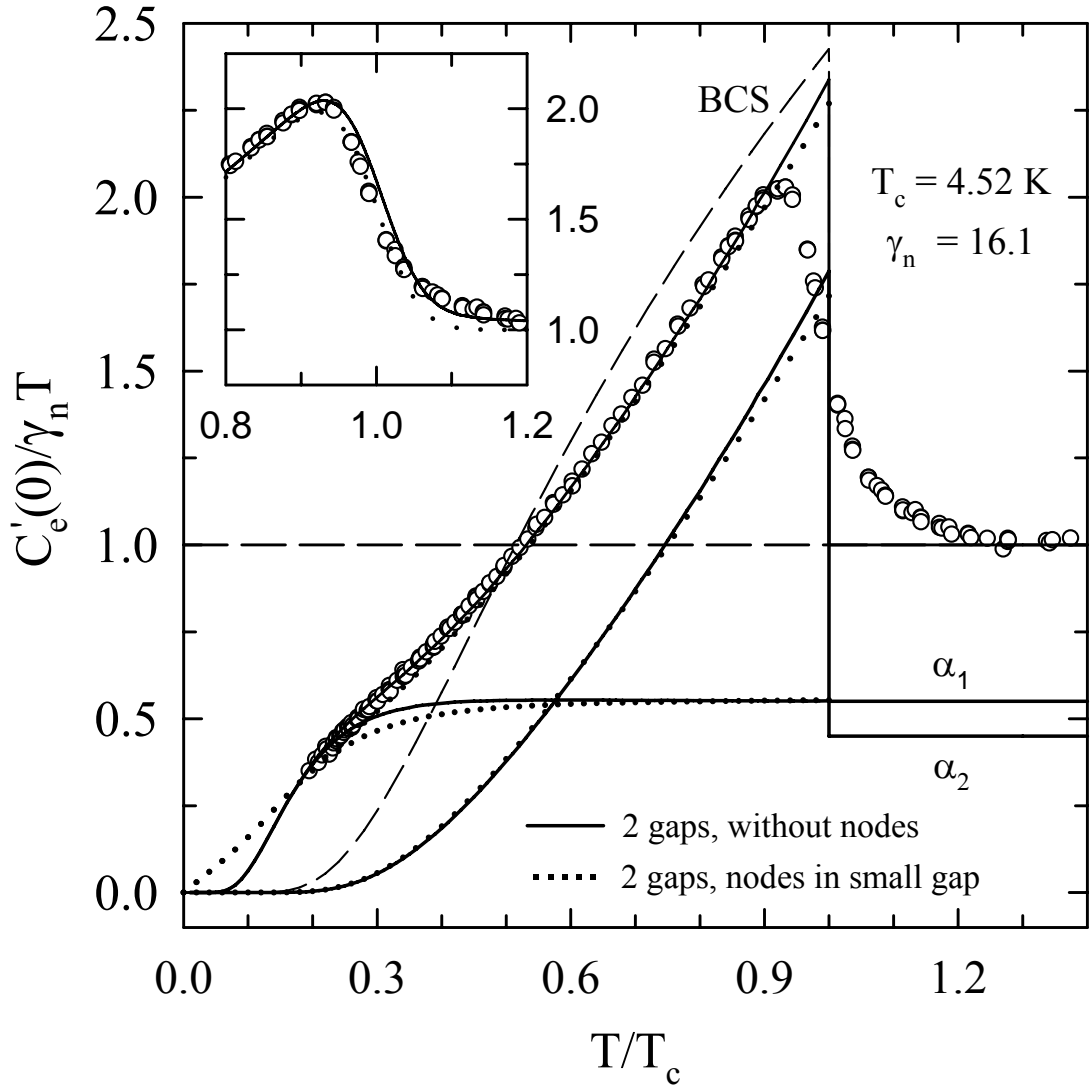

FIG. 2. Sample-2 electron specific heat compared with BCS theory and two 2-gap fits. Inset: Gaussian fits to the zero-field data near $T_c$, with (solid line) and without (dotted line) a fluctuation contribution.

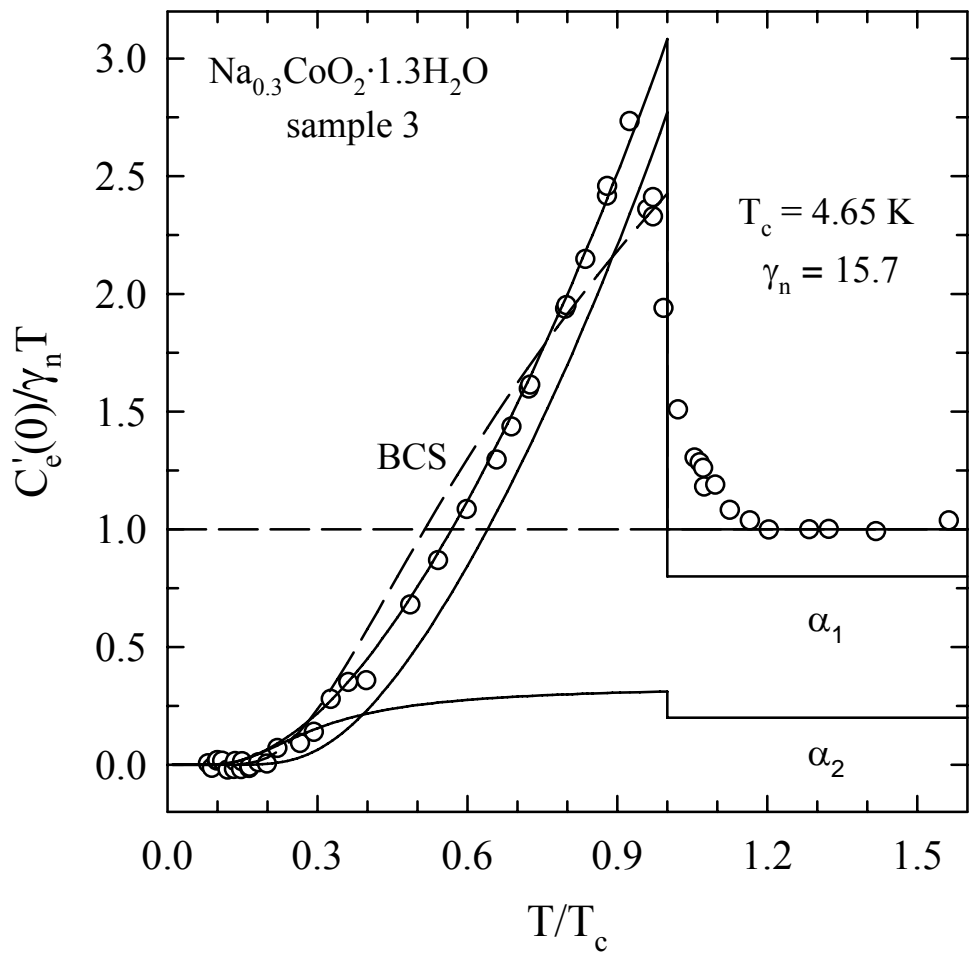

FIG. 3. Sample-3 electron specific heat compared with BCS theory and a 2-gap fit.

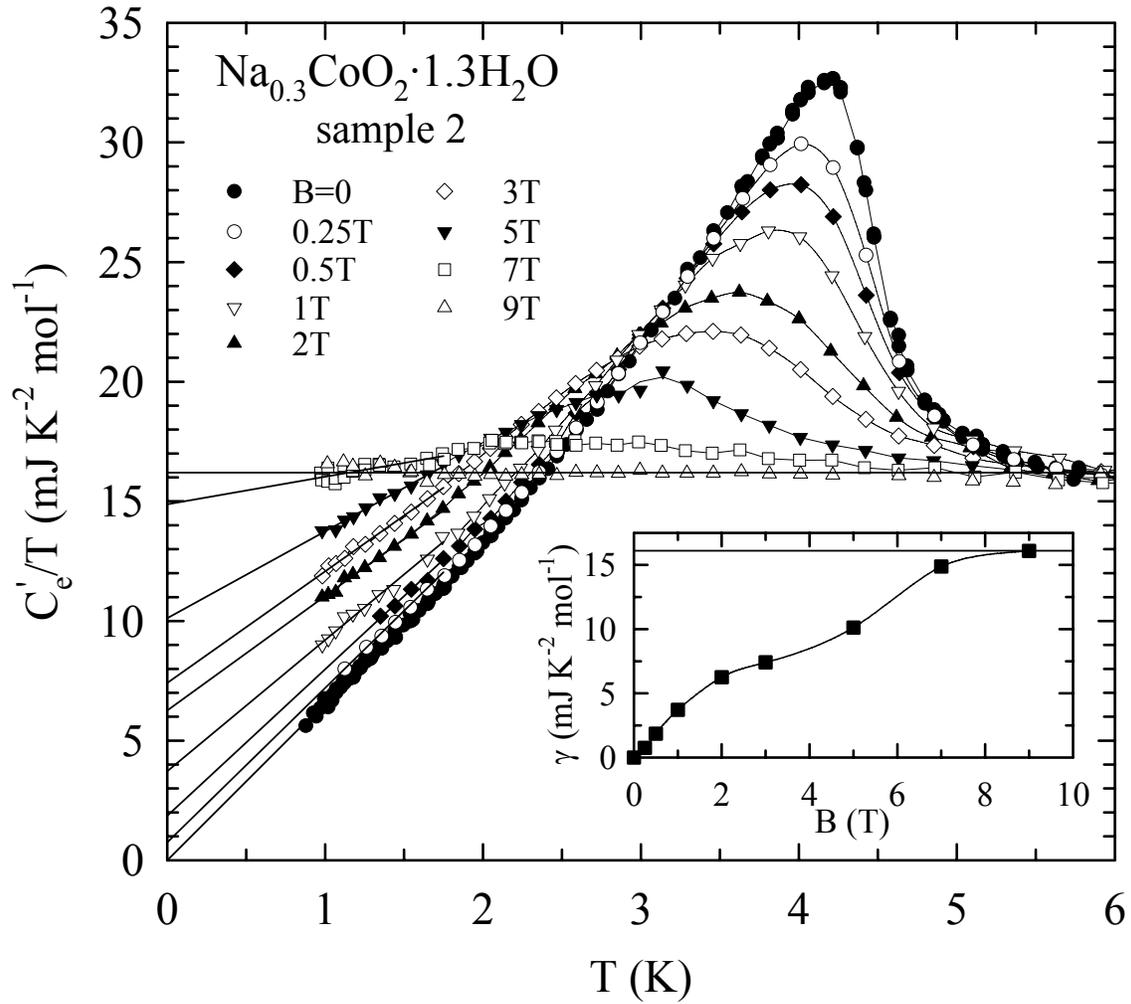

FIG. 4. Sample-2 mixed-state electron specific heat. Inset: $\gamma(B)$ vs B (see text).